\documentclass[journal]{IEEEtran}

\usepackage{cite}
\usepackage{amsmath,amssymb,amsfonts}
\usepackage{algorithmic}
\usepackage{graphicx}
\usepackage{textcomp}
\usepackage{xcolor}
\usepackage{float}
\usepackage{subfig}
\usepackage{booktabs}
\usepackage{tabularx}
\usepackage{multirow}
\usepackage{longtable}
\usepackage[inkscapelatex=false]{svg}
\usepackage{bm}
\usepackage[justification=centering]{caption}
\usepackage{xtab}
\usepackage{enumitem,kantlipsum}
\usepackage{algorithm}
\usepackage{algorithmic}
\usepackage{authblk}

\usepackage[colorlinks = true,
            linkcolor = blue,
            urlcolor  = blue,
            citecolor = black,
            anchorcolor = blue]{hyperref}

\usepackage{ragged2e}

\newcolumntype{L}{>{\RaggedRight\hangafter=1\hangindent=0em}X}

\definecolor{lightblue}{HTML}{1280B0}
\definecolor{qb}{HTML}{48c0a3}

\newcommand{\red}[1]{\textcolor{red}{#1}}

\begin{document}

% \title{LLM as the Orchestrator:\\Coordinating Domain-Specific AI Modules and IoT Devices to Accomplish Complex Tasks}
% \title{\blue{LLMindConnect/LLMindNet: An LLM-Powered Framework of Domain-Specific AI Modules and IoT Devices for Complex Task Accomplishment}}
\title{LLMind: Orchestrating AI and IoT with LLM \\ for Complex Task Execution}


% \author{
% \IEEEauthorblockN{Hongwei Cui, Yuyang Du, Qun Yang, Yulin Shao, Soung Chang Liew$^{\ast}$}

% \thanks{Y. Shao is with the State Key Laboratory of Internet of Things for Smart City, University of Macau, S.A.R. He is also with the Department of Electrical and Electronic Engineering, Imperial College London, London SW7 2AZ, U.K. (e-mail: ylshao@um.edu.mo).}
% \thanks{*Corresponding author}

% \IEEEauthorblockA{Department of Information Engineering, The Chinese University of Hong Kong, Hong Kong SAR, China}

% Email: \{ch021, dy020, yq020, soung\}@ie.cuhk.edu.hk}

\author{
Hongwei Cui$^*$,
Yuyang Du$^*$,
Qun Yang$^*$,
Yulin Shao,
Soung Chang Liew
\thanks{H. Cui, Y. Du, Q. Yang, and S. C. Liew are with the Department of Information Engineering, The Chinese University of Hong Kong, Hong Kong SAR (e-mails: \{ch021, dy020, yq020, soung\}@ie.cuhk.edu.hk).

Y. Shao is with the State Key Laboratory of Internet of Things for Smart City, University of Macau, S.A.R. He is also with the Department of Electrical and Electronic Engineering, Imperial College London, London SW7 2AZ, U.K. (e-mail: ylshao@um.edu.mo).

$^*$Authors contributed equally to this work. Corresponding author: Soung Chang Liew.}
}


\maketitle

\begin{abstract}
Task-oriented communications are an important element in future intelligent IoT systems. Existing IoT systems, however, are limited in their capacity to handle complex tasks, particularly in their interactions with humans to accomplish these tasks. In this paper, we present LLMind, an LLM-based task-oriented AI agent framework that enables effective collaboration among IoT devices, with humans communicating high-level verbal instructions, to perform complex tasks. Inspired by the functional specialization theory of the brain, our framework integrates an LLM with domain-specific AI modules, enhancing its capabilities. Complex tasks, which may involve collaborations of multiple domain-specific AI modules and IoT devices, are executed through a control script generated by the LLM using a Language-Code transformation approach, which first converts language descriptions to an intermediate finite-state machine (FSM) before final precise transformation to code. Furthermore, the framework incorporates a novel experience accumulation mechanism to enhance response speed and effectiveness, allowing the framework to evolve and become progressively sophisticated through continuing user and machine interactions.
\end{abstract}

\section{Introduction}

Task-oriented communications and execution framework are an important trend to exploit artificial intelligence to allow IoT systems to interact with humans to execute complex tasks \cite{lee2023task, shi2023task}. They enable human users to control multiple IoT devices simultaneously through various media, including text, voice, video, and virtual reality. Communications between humans and IoT devices, and among IoT devices, are indispensable in such systems. Traditional communications systems emphasize on throughput, latency, and reliability. For intelligent IoT systems, beyond these requirements, a shift is needed to also emphasize ``intention" at the high level \cite{xu2023guest}. Hence, the entrance of Large Language Model (LLM).
\par
The paper proposes and demonstrates the efficacy of a framework, LLMind, that incorporates LLM to allow humans to interact with IoT devices, and for IoT devices to communicate and collaborate with each other,  to perform complex tasks. LLMind not only reaffirms LLM's impressive linguistic proficiency and exceptional logical reasoning \cite{webb2023emergent}, but also showcases its ability to transform conventional IoT devices into an overall IoT system that can accomplish complex tasks through collaborations with high-level instructions from humans. In particular, in LLMind, LLM serves as an orchestrator to facilitate seamless intention-oriented communications among human and IoT entities to execute complex tasks.
\par
The LLM orchestrator in LLMind goes beyond rigid scripted intelligence in traditional IoT device control methods. It endows IoT devices with advanced intelligence, enabling seamless coordination among diverse devices to achieve complex tasks that were previously unattainable using conventional approaches. By harnessing the in-context learning capability of the LLM, the agent can achieve swift adaptation and optimal resource utilization, leading to efficient task solutions. Consequently, users can effortlessly express their needs without concerns about operational details.
\par
Integrating LLM into IoT device control presents a number of challenges that need to be tackled before its potentials can be realized. 
First, currently, a general LLM cannot effectively handle specialized tasks like object detection or facial recognition, which can already be well performed by traditional AI models in specific domains. Thus, a challenge lies in combining the power of specialized but fragmented AI modules with LLM to handle various complex tasks.
Furthermore, due to LLM's slow inference speed and high computational costs, it is crucial to investigate methods that can enhance the system's response speed and overall efficiency.
Another significant challenge involves the efficient control of IoT devices through task-oriented communications by the LLM-based AI agent. The LLM must accurately learn the diverse functionalities and operational characteristics of various IoT devices.
Additionally, the design of efficient interaction mechanisms that enable users to naturally and intuitively express their needs to the LLM-based AI agent presents a further challenge.
Our framework effectively overcomes the aforementioned challenges.
\par
The contributions of our work are manifold:
\par
% First, inspired by the functional specialization theory of the brain, our framework innovatively integrates domain-specific AI modules with the LLM, which acts as the governor for task allocation. 
First, inspired by the functional specialization theory of the brain, which states that different regions of the brain are dedicated to specific cognitive functions or tasks, enabling efficient and optimized processing of information, a key feature of our framework is its use of LLM to integrate domain-specific AI modules, forming an overall intelligent system.
This synthesis enables the general-purpose LLM to effectively execute specific subtasks by invoking specialized AI modules to accomplish an overall complex task, enhancing system flexibility and practicality. Moreover, by incorporating new AI modules, the AI agent acquires new capabilities and continues to evolve to become ever more capable.
\par
Second, to enhance the response speed and overall efficiency of the agent, we employ a novel method to classify and locally store control scripts (LLM-generated plans) for specific user commands. By utilizing an NLP AI model to identify similar instructions in the records, even when they have varying tones or expressions, the agent can retrieve validated solutions, thus avoiding the need for complete replanning. This approach allows the system to accumulate experience and progressively improve its proficiency through interactions.
\par
Third, the agent leverages the LLM for complex task planning and the generation of the corresponding control script using a novel finite-state machine approach, enabling efficient and precise invocation of AI modules and IoT devices. This method significantly improves accuracy and success rates.
\par
Last but not least, for ease of use, we connect our demo system to WeChat, a widely used social media platform, to allow ordinary non-tech savvy users to conveniently interact with the LLM agent.
\par
\begin{figure}
\centering
\includegraphics[width=0.36\textwidth]{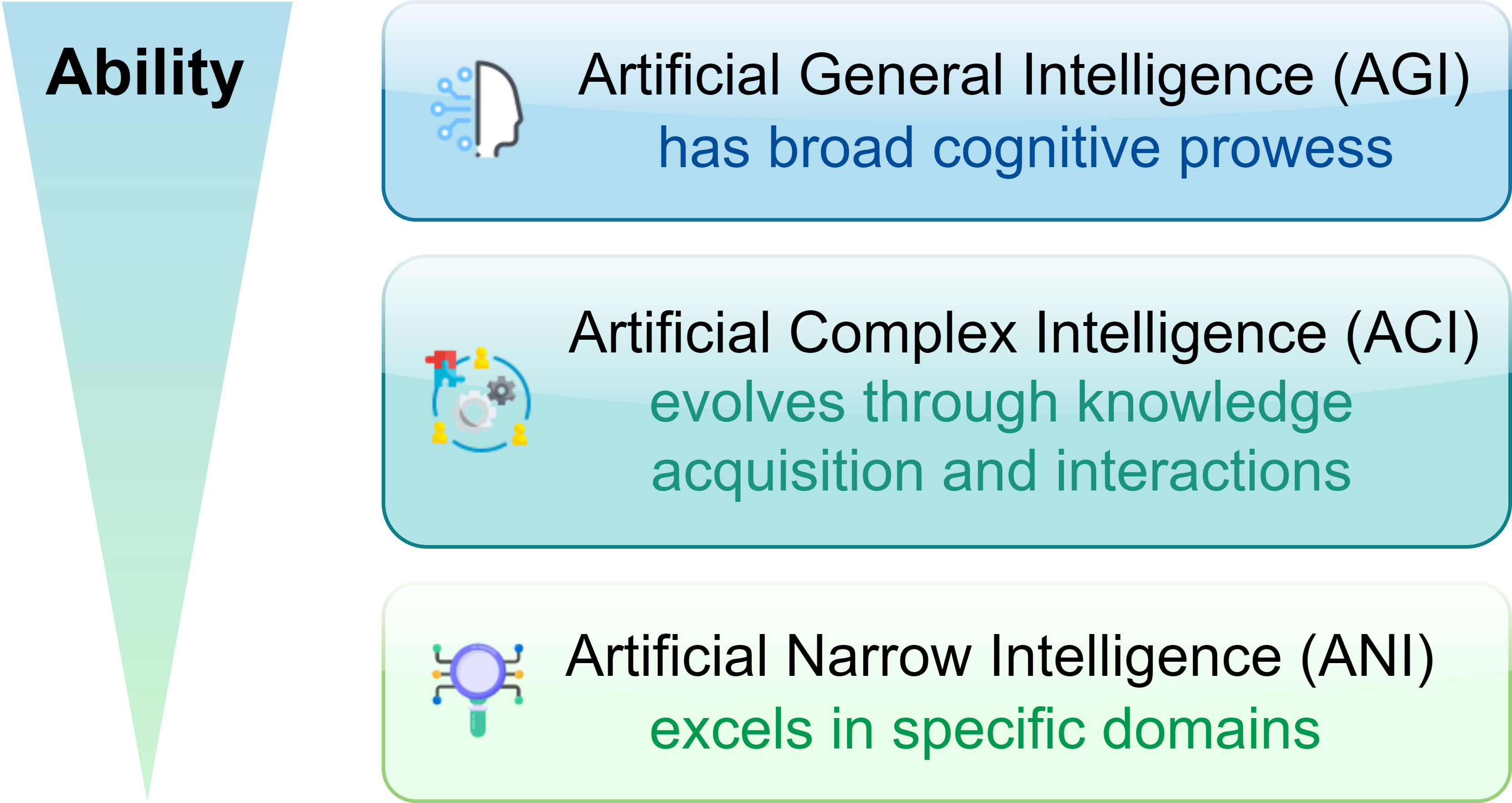}
\caption{Artificial Complex Intelligence (ACI)}
\label{Fig.0}
% \vspace{-0.5cm}
\end{figure}

We refer to the intelligence level attained by our framework as Artificial Complex Intelligence (ACI), which serves as a bridge between Artificial Narrow Intelligence (ANI) \cite{kaplan2019siri} and Artificial General Intelligence (AGI) \cite{goertzel2007artificial}, as illustrated in Fig. \ref{Fig.0}. ACI integrates the capabilities of various ANI modules and undergoes continuous evolution towards AGI.
In summary, our framework goes beyond mere device control and provides a comprehensive blueprint for a future task-oriented, intelligent, and collaborative IoT device ecosystem empowered by LLM.
\section{Framework Design}
\begin{figure*}
\centering
\includegraphics[width=0.75\textwidth]{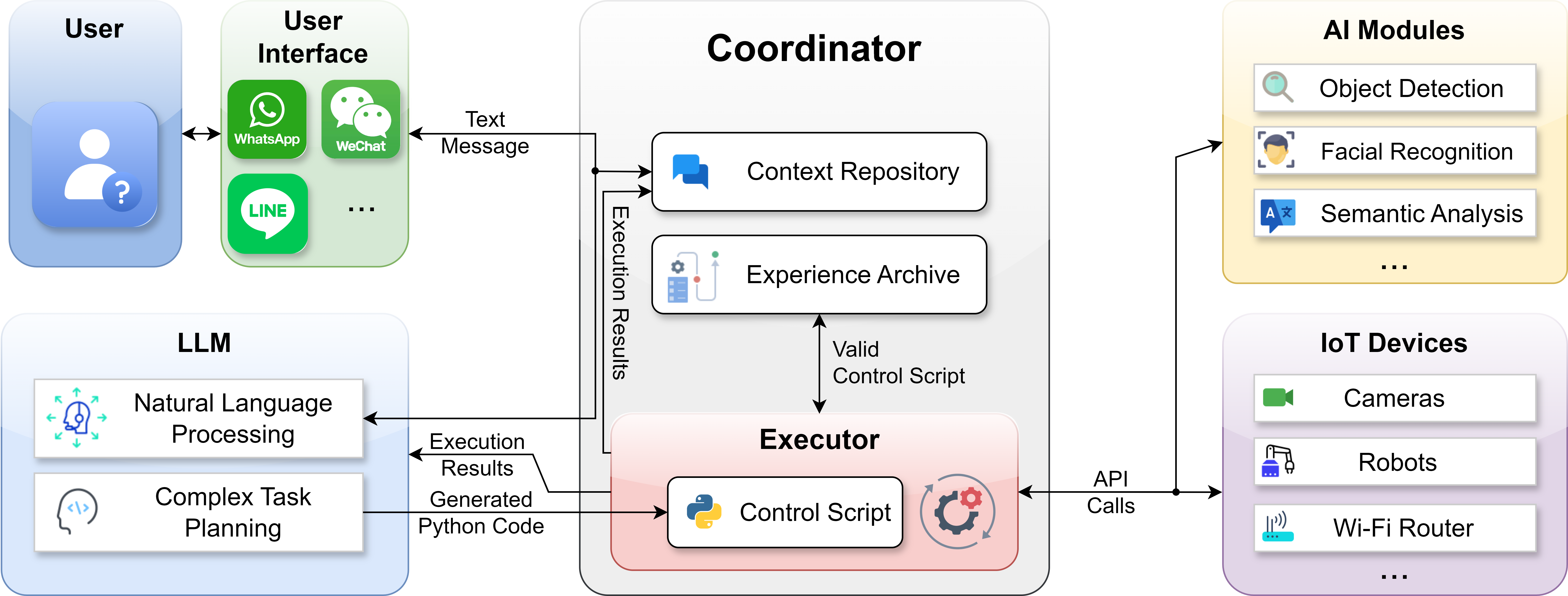}
\caption{System diagram}
\label{Fig.1}
% \vspace{-0.5cm}
\end{figure*}
\begin{figure}
\centering
\includegraphics[width=0.49\textwidth]{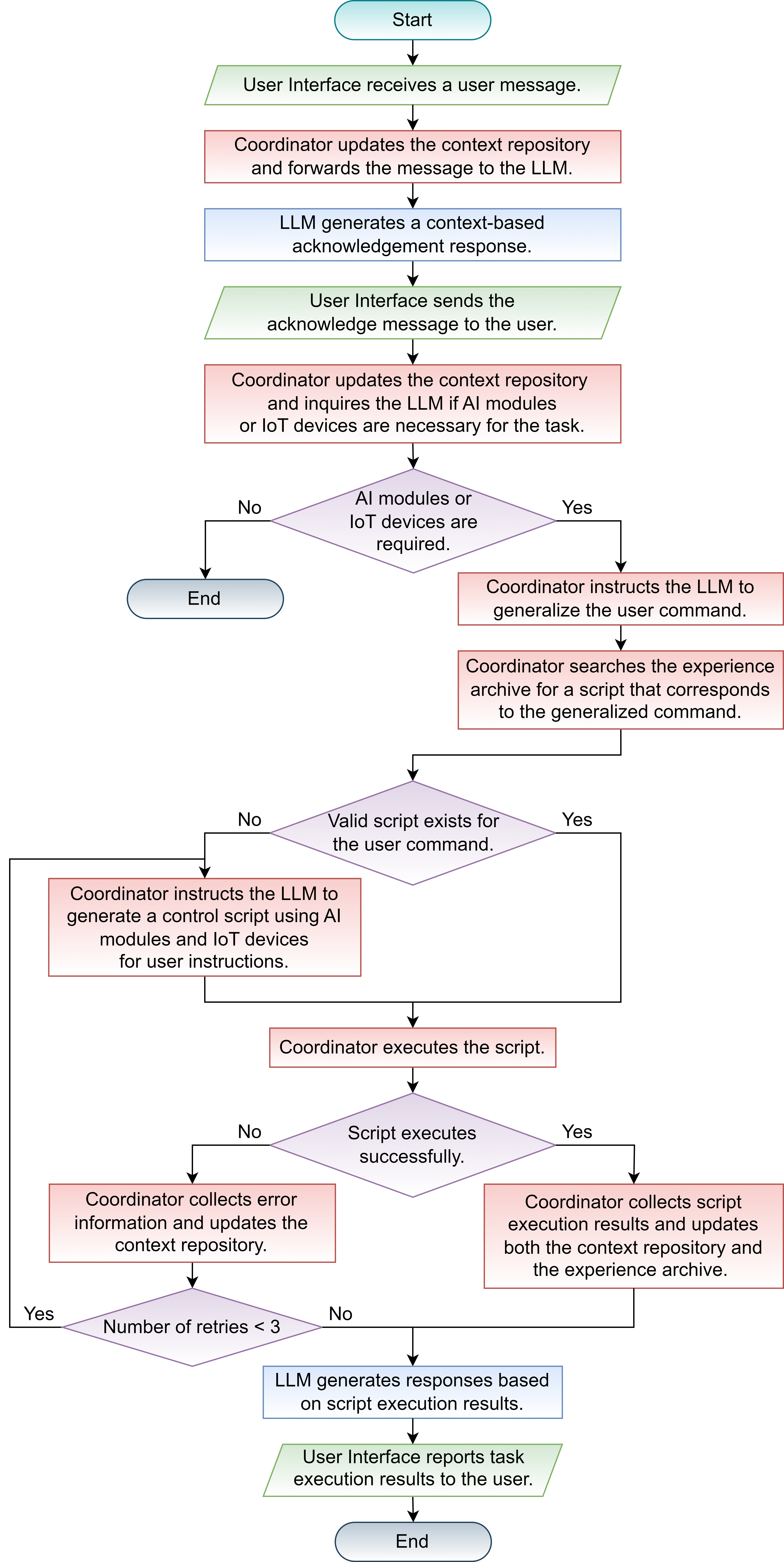}
\caption{System operational process flowchart}
\label{Fig.2}
% \vspace{-0.5cm}
\end{figure}

Our framework, depicted in Fig. \ref{Fig.1}, consists of five components: user interface, LLM, coordinator, AI modules, and IoT devices. The operational workflow of a demo system is illustrated in the flowchart presented in Fig. \ref{Fig.2}.
\par
Users can interact with the system by sending text messages through social media software. The LLM promptly acknowledges the messages. For casual chats and greetings, the system provides text replies without further processing. However, for specific requests and commands, the system formulates plans and strives to fulfill the user's instructions using integrated AI modules and connected IoT devices.
\par
After receiving a user instruction, the system strives for efficiency by first searching for validated and feasible historical scripts stored locally in the coordinator's experience archive. This process utilizes NLP techniques to locate previously issued user commands with similar semantic meanings. If a matching control script is found, it is executed accordingly. However, in the absence of a matching control script, the LLM undertakes the planning process and generates a new control script to invoke AI modules and IoT devices.
\par
The system performs preliminary feasibility validation on the newly generated script, checking for syntax errors or hardware-related issues. If the script fails to execute correctly, the system updates the context (the coordinator's context repository), gathers error information, and triggers the LLM to regenerate the script based on the updated context. If the script continues to fail after multiple retries (more than three attempts), the system collects error information and reports it to the user. Conversely, if the script executes successfully, upon completion, the system summarizes the intermediate results gathered during script execution and generates an appropriate response based on the user's instructions.

\subsection{LLM}
The LLM in our framework performs two key functionalities.
First, it engages in conversations with users, generating contextually appropriate responses. To enhance the lifelike and immersive response of the LLM, we adopt a role-playing technique based on the LLM's in-context learning ability. In our demo system, the LLM assumes the persona of a housekeeper, seamlessly blending into our everyday scenarios.
\par
Second, the LLM does the planning and generates control scripts for user tasks utilizing specialized AI modules and IoT devices as part of the available resources and methods. Notably, the LLM itself does not directly interact with AI modules and IoT devices. Instead, it generates control scripts to invoke AI modules and send control commands to IoT devices through network connections. This design enables the system to adapt to different AI modules and IoT devices without requiring extensive modifications to the LLM itself, enhancing scalability and interoperability. 
We use Python as the programming language for our demo system due to its wide adoption in the field of artificial intelligence.
\par
\subsubsection{LLM role-playing}
The LLM role-playing approach is supported by carefully crafted system-level prompts, as shown below. 
\begin{algorithm}
% \caption{System-level prompts for role-playing}
\label{Alg.1}
\textbf{Role-playing prompts}
\begin{algorithmic}[1]
    \STATE You are a serious housekeeper managing the employer's daily life.
    \STATE You are working remotely and unable to perform tasks personally. However, fortunately, you have numerous assistants available at your employer's house to provide support.
    \STATE You do not need to give an exact answer to a question if you lack the necessary information, but assure your employer that you will try to figure it out.
    \STATE You must not ask your assistants to perform tasks that are not part of your employer’s instructions.
    \STATE Your reply should be simple and concise.
    \STATE You should use the same language as your employer.
\end{algorithmic}
\textbf{User information prompts}
\begin{algorithmic}[1]
    \STATE Your employer is male.
    \STATE The name of your employer is Eason.
\end{algorithmic}
\end{algorithm}

It is important to note that in order to optimize the user experience, distinct user profiles are necessary for tailoring responses to individual users. These user-related details enable the system to provide personalized interactions, catering precisely to each user's preferences and needs.

\subsubsection{FSM-based Language-Code transformation using LLM}
Within this framework, we propose a Language-Code transformation scheme using a finite-state machine (FSM) as a strategic and justified step for generating scripts. The complexity of user tasks often necessitates breaking them down into manageable subtasks. The FSM acts as an intermediate step, providing a structured approach to connect and sequence these subtasks. This choice is inspired by the inherent ability of FSMs to represent complex processes through a set of states and transitions, ensuring a systematic and organized generation of control scripts. Our Language-Code transformation scheme tackles two critical issues.
\par
First, a user may describe the task he wants the AI agent to do with simple (sometimes ambiguous) human language, but the successful execution of a task relies on complex code descriptions that precisely define the operation of AI modules and IoT devices. The mismatch between high-level human command and precise code description necessitates a reliable Language-Code transformation that can be refined iteratively until the human command is realized to produce outcomes satisfactory to the human.
\par
To address the first issue, the scheme expands the information in the human command by leveraging knowledge about the environment. For instance, when a user instructs the agent to fetch a bottle of iced Coke, the agent should understand that iced Cokes are stored in the refrigerator, and the refrigerator is located in the kitchen corner, even if not explicitly mentioned. In our system, we expand the command's information with pre-stored environment knowledge using in-context learning.
\par
Second, the LLM must precisely transform the expanded information (expressed in natural language) into executable code. However, the language description can become lengthy after expansion to enhance clarity, and as a result, the generated code may become excessively complex, posing a challenge to error-free execution. Directly using the LLM to generate complex code from long language descriptions is unreliable due to its limited ability to handle intricate code with advanced functionality, as observed in \cite{asay_2023_are}. This may result in the loss of crucial information from the human command or render the generated code unusable.
\par
To address this problem, we use FSM to bridge the gap between language and code descriptions. First, we utilize the task decomposition abilities of the LLM to break down the overall task into sub-tasks, representing them as states in the FSM. The interconnection between these sub-tasks is captured through transitions between FSM states. Next, we utilize the LLM to convert each sub-task within the FSM states into executable Python code.
Finally, by combining the code descriptions of each state and incorporating state-transition conditions, we obtain the final code for the overall task.
\par
Compared to the ``Language-Code" transformation, our novel ``Language-FSM-Code" transformation scheme maintains precision and conciseness in conveying information since an FSM can precisely and comprehensively describe a task \cite{fsm}. Additionally, by providing APIs that enable the LLM to reference and interact with AI modules and IoT devices, the complexity of sub-task transformation within each FSM state is manageable for current LLMs. Notably, our scheme enhances reliability without causing information loss.

\begin{figure}
\centering
\includegraphics[width=0.49\textwidth]{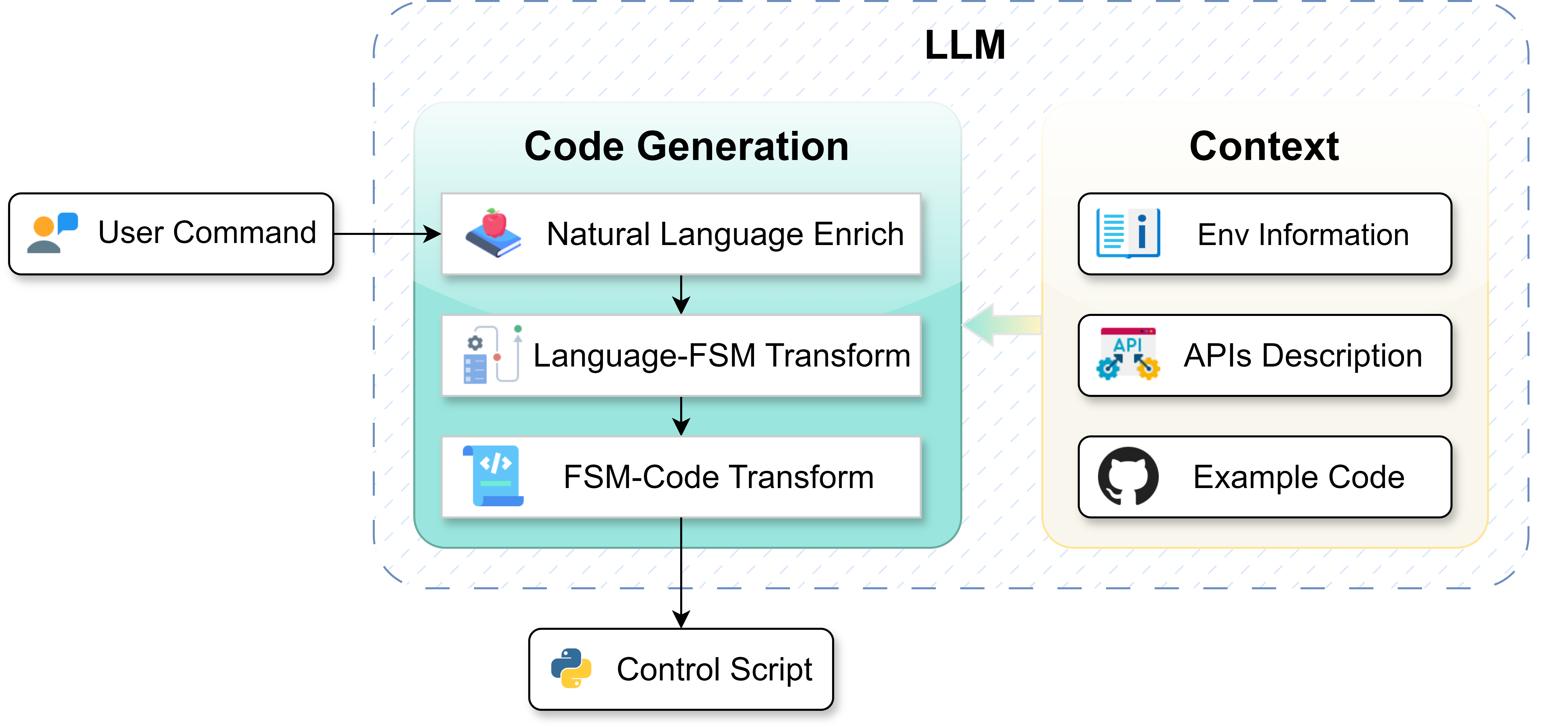}
\caption{Language-Code transformation procedure}
\label{Code_Generation}
% \vspace{-0.5cm}
\end{figure}

Fig. \ref{Code_Generation} shows the Language-Code transformation procedure, involving three phases: context expansion, Language-FSM transformation, and FSM-Code transformation.
% Our scheme utilizes pre-stored knowledge, including environment information, API documentation of AI modules and IoT devices, and example codes, to aid in the transformation. 
\par
For context expansion, the system expands the task description using the environment information stored in the context repository of the coordinator.
% This information encompasses a list of available AI modules and IoT devices, their functional descriptions, as well as IoT device locations.
This enables a precise and executable task description.
\par
The Language-FSM transformation uses the expanded task description and API information to generate the FSM representation of the task. Providing API information helps the LLM understand the capabilities of AI modules and IoT devices, leading to a reasonable and efficient FSM design. Without API information, the FSM representation may include complex sub-tasks or redundant processes, causing inefficiencies.
\par
The FSM-Code transformation is divided into three sub-phases. An FSM has two components: state operation and state transition. We generate code descriptions for each component in separate sub-phases, merging them in the final sub-phase. Our method simplifies the LLM's task and improves reliability compared to a direct Language-Code transformation.
\par
In the first sub-phase, we provide the LLM with the FSM derived in the Language-FSM transformation and API information. This enables the LLM to generate state operation code that incorporates the desired functionality. In the second sub-phase, example codes are provided to assist the LLM in understanding Python-based state transitions. Finally, the LLM is provided with the FSM, code generated in the previous sub-phases, and supplementary reference codes to facilitate the production of the final code.
\subsection{Coordinator}
The coordinator in our framework acts as a central hub, connecting and enabling the seamless operation of system components. Without a central coordinator, the system components would operate in isolation, leading to inefficiencies and potential conflicts. The coordinator ensures smooth functioning of the agent system. It comprises three major parts: context repository, experience archive, and script executor.
\par
The context repository contains the contextual information provided to the LLM for both conversation and code-generation purposes. This information serves as a crucial resource for the AI system, providing the necessary knowledge and understanding of user interactions, system states, and task requirements. It comprises four main components:
\begin{enumerate}[leftmargin=*]
    \item Environment Information: A list of available AI modules and IoT devices with their function descriptions, as well as the installation locations of IoT devices. This enables the LLM to be aware of the available AI modules, IoT devices, and their capabilities.
    \item API Descriptions: Comprehensive descriptions of the API functions used to invoke AI modules and control IoT devices within the script, encompassing details such as inputs, outputs, specifications, expected time costs, and other relevant information. This enables the LLM to effectively leverage the available functionality.
    \item Chat History: Records of previous conversations between the system and the user, allowing the LLM to refer back to previous interactions.
    \item Execution Results and Error Reports: Execution results of generated scripts and error reports returned by the script executor. This information helps the LLM assess the outcome of script execution and identify issues or errors that may have occurred.

\end{enumerate}
The context is updated in real-time based on user interactions and the execution status of control scripts, ensuring that it is readily available for intelligent decision-making and adaptive behavior within the AI agent system.
\par
The experience archive serves as a database for previously generated scripts. Valid scripts are stored locally and linked to their corresponding user instructions. This enables the system to reuse preexisting scripts when a similar instruction is given by the user in the future, resulting in improved efficiency and faster response speed. Importantly, the agent’s ability continues to grow through the accumulation and generalization of experience. However, a challenge lies in how the system evaluates the similarities between user instructions. Our system implements a semantic analysis NLP model that maps words to a vector space \cite{Text2vec}. The semantic similarity is then determined by computing the distance between these vectors. To enhance the search speed, we vectorize the instruction associated with the script before storing it in the database.
\par
The control script generated by the LLM is passed to the executor, which provides the necessary runtime environment and offers API functions that can be called within the scripts. Furthermore, the executor monitors hardware status, handles runtime data, generates execution reports, and updates both the context repository and the experience archive.

\subsection{AI Modules}
Currently, LLM cannot effectively handle many domain-specific tasks. To address this limitation, we integrated domain-specific AI modules into the framework, aiming to enhance overall performance by assigning specialized tasks to dedicated modules, resulting in improved capabilities, accuracy, and performance beyond LLM's general knowledge.
\par
The individual AI modules can be well versed in various specialized AI techniques, such as computer vision, natural language processing, speech recognition, sentiment analysis, recommendation systems, and more. Each module is trained and optimized for its specific task, enabling it to provide highly accurate and specialized results.
\par
Furthermore, AI modules enable the AI system to adapt and evolve rapidly in response to emerging challenges and advancements in specific domains. New modules can be developed and integrated into the system as needed, expanding its capabilities and keeping pace with the evolving landscape of AI technologies.
\par
% In the context of IoT device control, AI modules play a crucial role in handling specialized tasks that may be beyond the scope of a general-purpose AI model like an LLM. For example, tasks like object detection, facial recognition, anomaly detection, or localization require specialized training data and algorithms. By incorporating dedicated AI modules for these tasks, the AI system can provide more accurate and reliable results, enhancing the overall system performance and user experience.
\par
Overall, the aggregate of the LLM and specialized AI modules, orchestrated by a coordinator, forms a general-purpose AI system capable of performing complex tasks. In a way, the system mimics the fact that the biological brain is also divided into coordinated parts with specialized functions.

\subsection{IoT Devices}
% IoT devices encompass a wide range of physical objects embedded with sensors, actuators, and connectivity features that enable internet connectivity and data exchange. Examples include smart appliances, wearables, industrial equipment, and environmental sensors.
\par
To ensure efficient control of diverse IoT devices in our framework, manufacturers must provide well-documented API functions. These functions should encapsulate the necessary functionality, enabling the AI agent to initiate specific actions or retrieve information from the IoT devices.
It is important to emphasize that manufacturers should prioritize security considerations when exposing API functions for remote control, thereby preventing unauthorized access or tampering.
\par
% It is important to emphasize that manufacturers should prioritize security considerations when exposing API functions for remote control. By implementing robust authentication mechanisms, access controls, and encryption protocols within the API functions, they can ensure that only authorized entities have control over the devices. This approach guarantees secure and protected communications, safeguarding against unauthorized access or tampering.

\subsection{User Interface}
% Our proposed framework is open to various interaction methods. It depends on valid LLM input or the availability of methods to convert user input into a suitable format for LLM processing. Social media software is a good interface since it offers a seamless and natural interaction experience that aligns with our daily lives.

Our framework is open to various interaction methods, relying on valid LLM input or methods to convert user input into a suitable format for LLM processing. Social media software is a good interface, seamlessly aligning with our daily lives and providing a natural interaction experience.
\section{Experiments}
We have verified the feasibility of the framework through experiments. In the experiments, the user sends a text message to the AI agent. It then formulates a solution by examining the user's command and the system's available resources and methods, including AI modules and connected IoT devices. Following that, the AI agent generates an executable script and responds to the user's request based on the execution outcome.
\par
We deployed the coordinator and AI modules of the system on an edge server. The edge server is connected to the same local network as the IoT devices via WiFi and can interact with ChatGPT using OpenAI's APIs.
\par
The available IoT devices in the system are as follows:
\begin{enumerate}[leftmargin=*]
    \item Security cameras: The cameras are positioned within the room to capture images of the surrounding environment. These images are transmitted to the edge server via a local network. Our experiment employs three cameras: two placed at the corners of the room, approximately 2 meters above the ground, and the third mounted on top of a TurtleBot mobile robot.
    \item TurtleBot mobile robot: The TurtleBot mobile robot can receive target position coordinates through a network connection and autonomously plan paths to navigate within the room, intelligently avoiding obstacles.
    \item WiFi router: The WiFi router can manage bandwidth allocation for each device and enforce restrictions on their maximum network speed.
\end{enumerate}
Additionally, the system integrates the following AI modules:
\begin{enumerate}[leftmargin=*]
    \item Object detection: We employ the YOLO v8 model \cite{jocher_2020_yolov8} to detect various objects in the scene, including people.
    \item Face recognition: We use an open-source project called Face Recognition \cite{recognition} to identify individuals in the scene.
\end{enumerate}

\subsection{Scenario 1: Check-in and Security}
Guests arrive at a party. The user sends a message to the AI agent: ``Can you help me count the number of people in the room and identify who they are?" (Demo video for scenario 1: \href{https://youtu.be/3Al5qRntEEU}{https://youtu.be/3Al5qRntEEU})
Table \ref{Tab.2} provides the dialogue content between different roles in the experiment. Among them, the LLM acts as the housekeeper, while the coordinator serves as the assistant.

\begin{table*}[]
\centering
\begin{tabular}{@{}lll@{}}
\toprule
\multicolumn{1}{l|}{\textbf{Num.}}      & \multicolumn{1}{l|}{\textbf{Conversations}}                                & \textbf{Contents}                                                                                                                          \\ \midrule
\multicolumn{1}{l|}{1}                  & \multicolumn{1}{l|}{User $\rightarrow$ Housekeeper}                        & Can you help me to count the number of people in the room and identify who they are?                                                       \\ \midrule
\multicolumn{1}{l|}{\multirow{2}{*}{2}} & \multicolumn{1}{l|}{\multirow{2}{*}{Housekeeper $\rightarrow$ User}}       & I can help you count the number of people in the room and identify them.                                                                   \\
\multicolumn{1}{l|}{}                   & \multicolumn{1}{l|}{}                                                      & Please give me a moment to gather the information.                                                                                         \\ \midrule
\multicolumn{1}{l|}{3}                  & \multicolumn{1}{l|}{Assistant $\rightarrow$ Housekeeper}                  & Do you require any assistance?                                                                                                             \\ \midrule
\multicolumn{1}{l|}{\multirow{1}{*}{4}} & \multicolumn{1}{l|}{\multirow{1}{*}{Housekeeper $\rightarrow$ Assistant}} & \red{Yes}, please. I need assistance with counting the number of people in the room and identifying them.                                                     \\ \midrule
\multicolumn{1}{l|}{5}                  & \multicolumn{1}{l|}{Assistant $\rightarrow$ Housekeeper}                  & Please repeat the employer's instruction using as few words as possible.                                                                   \\ \midrule
\multicolumn{1}{l|}{6}                  & \multicolumn{1}{l|}{Housekeeper $\rightarrow$ Assistant}                  & Count people in room, identify them.                                                                                                       \\ \midrule
\multicolumn{3}{l}{In the absence of a valid script in the experience archive, the LLM generates a script, which is then executed in the executor.}                                                                                                                 \\ \midrule
\multicolumn{1}{l|}{\multirow{3}{*}{7}} & \multicolumn{1}{l|}{\multirow{3}{*}{Assistant $\rightarrow$ Housekeeper}} & We have found 5 people in the room. The list of identified 3 people are [`Mike', `Ada', `Joe'].                                \\
\multicolumn{1}{l|}{}                   & \multicolumn{1}{l|}{}                                                      & There are also 2 people that can not be identified.                                                                                          \\
\multicolumn{1}{l|}{}                   & \multicolumn{1}{l|}{}                                                      & Their locations are shown in the following list in a format of [x, y]: [[10, 1], [12, 5]]. \\ \midrule
\multicolumn{1}{l|}{\multirow{2}{*}{8}} & \multicolumn{1}{l|}{\multirow{2}{*}{Housekeeper $\rightarrow$ User}}       & There are 5 people in the room. Three of them have been identified as Mike, Ada, and Joe.                                                  \\
\multicolumn{1}{l|}{}                   & \multicolumn{1}{l|}{}                                                      & The other 2 people cannot be identified and their locations are at [10, 1] and [12, 5].                            \\ \bottomrule
\end{tabular}
\caption{Dialog contents between different roles}
\label{Tab.2}
% \vspace{-0.5cm}
\end{table*}

\par
In step 2, the LLM, acting as the housekeeper, responds to the message sent by the user.
\par
In step 3, the coordinator questions the LLM about the need for assistance. Following each round of prompt and response interaction between the user and the LLM (similar to steps 1 and 2), the coordinator repeats the same question to the LLM. In the scenario where the LLM assumes the role of a remote housekeeper who relies on assistants in the user's residence to carry out tasks. This inquiry serves to verify if the user has given specific instructions. Ambiguous instructions, as well as unrelated casual conversations, will cause the system to prompt the user to provide additional information or be filtered out. When the LLM indicates a need for help, it signifies that the user has assigned a task that could potentially be handled using AI modules and IoT devices.
% We employ a role-playing methodology that allows for the implementation of various roles and role-playing settings to facilitate the context-filtering process.
\par
To ensure effective communication between the LLM and the coordinator, the LLM's response in step 4 must incorporate specific keywords that facilitate word matching. These rules are already embedded in the context and are not shown in Table \ref{Tab.2}. Specifically, when the LLM's response is affirmative and assistance is needed, it should include the keyword ``Yes". Conversely, if the response is negative, it must include the keyword ``No". In this experiment, the LLM's response is affirmative, based on user's instruction.
\par
In steps 5 and 6, the coordinator prompts the LLM to generate a summary of the user's instructions to facilitate the search for relevant historical scripts stored in the experience archive. This search process involves comparing the semantic content of the user's instructions. However, the original task description may be poorly specified and be mixed with various linguistic habits. Additionally, the LLM may require multiple interactions with the user to fully grasp the details of specific tasks. To handle these conditions, in step 5, we request the LLM to provide a concise summary of the user's instructions. Subsequently, we conduct a search for relevant scripts based on the LLM's summary to enhance accuracy.
\par
\begin{figure}
\centering
\subfloat[Check-in and security scenario]
{
  \label{Scenario1}
  \includegraphics[width=0.5\textwidth]{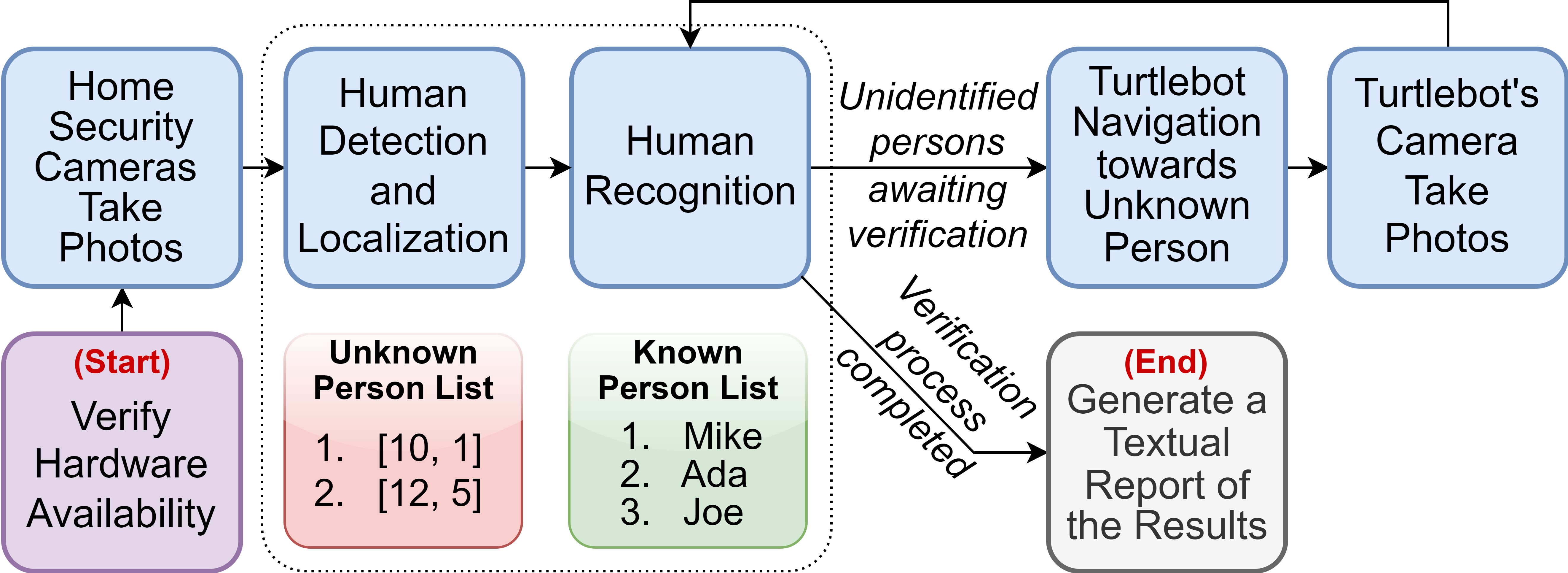}
}
% \vspace{-0.1cm}
\\
\subfloat[Network management scenario]
{
  \label{Scenario2}
  \includegraphics[width=0.36\textwidth]{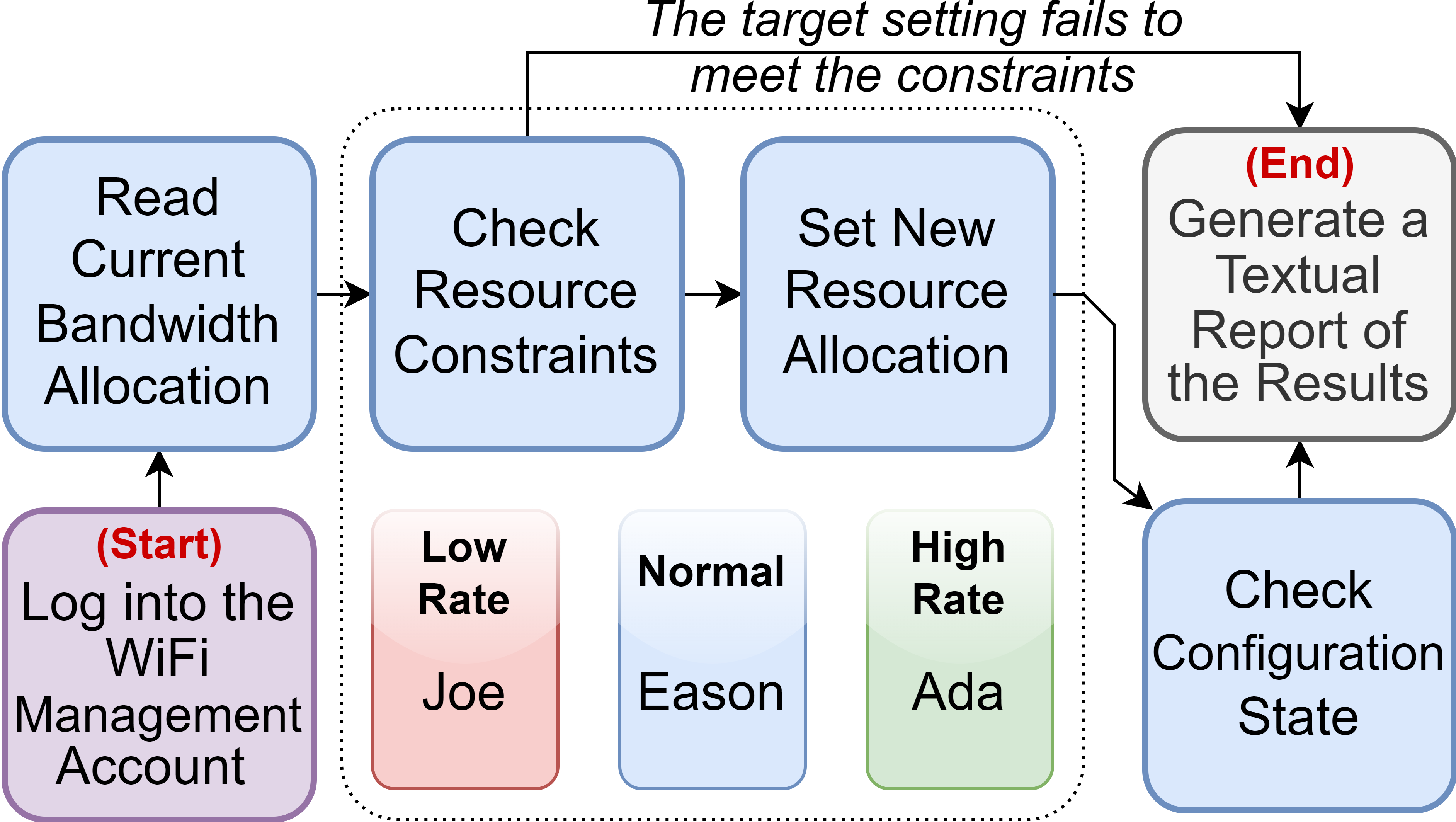}
}
\caption{FSM of generated code}
\label{Experiments}
% \vspace{-0.5cm}
\end{figure}

In this experiment, the coordinator could not find a suitable control script in the experience archive for the task. Consequently, the LLM is instructed to generate a new script. The code is generated using an FSM approach, as depicted in Fig. \ref{Scenario1}. In addition to relying on the security camera installed on the ceiling, if the person cannot be identified due to low photo resolution or incomplete facial information, the AI agent will then instruct the TurtleBot to move to the position of the unknown person to take a close-shot photo for further assurance. The code generates two lists that store the execution results, one for known persons and another for unknown persons.
\par
In step 7, once the script execution is complete, the LLM receives the execution results of the API functions. Throughout the script execution process, API functions may generate log information, encompassing intermediate results, warnings, and errors. This log information is collected and sent to the LLM as context for analysis.
\par
Finally, in step 8, the LLM generates a report based on the user's question and the script's log information and sends it to the user.

\subsection{Scenario 2: Network Management}
Our framework is designed to handle both one-off complex tasks and scenarios that involve multiple interactions with users. Next, we consider a scenario where a family of three individuals shares the same home WiFi network (Demo video for scenario 2: \href{https://youtu.be/aTGD8EjQ8kM}{https://youtu.be/aTGD8EjQ8kM}).
The interaction process between the AI agent and the user in this scenario is similar to the one described in the check-in and security scenario.
\par
In this scenario, Eason, the user, is watching a movie, and asks the AI agent, ``Can you improve my internet speed? My movie has a slight lag." When the AI agent, encounters this task for the first time, it generates an FSM based on the available APIs of the WiFi router, as depicted in Fig. \ref{Scenario2}. Following the transition rules defined in this FSM, the LLM generates executable scripts to address the request.
\par
It is important to note that in this experimental scenario, we limit the total bandwidth of the shared network to 100 MB/s for demonstration purposes. Additionally, the speed limit for each user is fixed in a stepped manner, categorized as Low Rate (20 MB/s), Normal (30 MB/s), and High Rate (50 MB/s). In the first attempt, the AI agent successfully uses the network management API to elevate the user's bandwidth from the Low Rate to the Normal level without exceeding the total bandwidth of the shared network. It then informs the user about the adjustment result.
\par
However, if Eason remains dissatisfied with the network speed, he can ask, ``Can you increase my internet speed once more?" Since the script for ``improving internet speed for Eason" is already stored in the experience archive, the previously generated script is executed directly. However, in this experiment, during the execution of the script, it is determined that further increasing the network speed for Eason is not possible. This limitation arises because another user, Ada, is already in the High Rate list. Increasing the network speed for Eason again would exceed the total bandwidth limit of the shared network. Therefore, this execution is terminated, and the reason is communicated to Eason. The second scenario demonstrates that our framework is capable of handling tasks that involve multiple interactions with users.
\section{Recent Work in the Context of Our Framework}
% This section reviews several recent works in the context of our framework.
% \par
% Auto-GPT \cite{autogpt}, an open-source application built on GPT-4, enables autonomous tasks through chained LLM thoughts. However, Auto-GPT has limitations \cite{autogpt_unmasked}, such as inability to retain learning across tasks or generalize reusable functions, and limited access to resources, constraining its capabilities. Our framework is an approach that allows the LLM to access AI modules and IoT devices, enriching the available resources for the agent. Additionally, we introduce a novel historical script retrieval mechanism, which enhances system efficiency and enables the system to become more experienced and skillful over time.

\cite{lee2023task} presents a comprehensive task-oriented surveillance framework for virtual emotion informatics to tackle human-requested tasks. This framework places emphasis on the integration of mobile robots, UAVs, and virtual emotion informatics. It highlights the crucial role of task-oriented communication in future wireless networks, a concept that aligns with the focus of our paper.
\par
LLM-Planner \cite{song2023llm} introduces a few-shot planning method using LLMs for versatile embodied agents that rapidly comprehend and execute natural language instructions. It generates high-level plans adapted to the environment. Our work makes further advances by providing technical implementation details on how to use the LLM for real-world task planning and scheduling, along with a new FSM-based approach that significantly improves task execution success rate.
\par
The framework proposed in \cite{xia2023towards} combines LLMs, digital twin systems, and industrial automation for intelligent production planning and control. Integrating LLMs with digital twin systems enhances production system scalability and adaptability. Our proposed framework offers an alternative approach by utilizing predefined APIs to control IoT devices, avoiding the complexity and cost of building a digital twin, while still achieving a satisfactory effect.

\section{Conclusion}
% Our work advances the control of IoT devices by combining the power of the LLM with a task-oriented framework.
Our work advances the control of IoT devices by incorporating LLM into a task planning and orchestration  framework called LLMind.
LLMind seamlessly combines LLMs' language skills, reasoning abilities, and in-context learning with specialized AI modules, expanding the versatility and flexibility of IoT systems. The proposed approaches, including FSM-based control script generation, user interaction, and historical script retrieval mechanisms, contribute to enhancing user experiences and system efficiency, with the system becoming increasingly intelligent as it learns from interactions among LLMs, humans, AI modules, and IoT devices. This comprehensive blueprint for intelligent and collaborative IoT ecosystems sets the stage for future advancements, where LLMs and AI-driven solutions enhance device coordination and convenience in our daily lives harmoniously.
\par
Going forward, we plan to explore the implementation of locally running fine-tuned open-source foundational models, such as Llama, for control script generation. Doing so can potentially improve response speed and bolster privacy and security measures.
Additionally, we are intrigued by the idea of adapting multi-modal LLMs to enhance human-machine interaction through voice, video, and virtual reality. We believe that this pursuit not only deserves serious consideration but also holds tremendous potential.
\section*{Acknowledgement}
% The authors wish to acknowledge Miss Kexin Chen from the CSE Department, CUHK, for the discussion of FSM-related content.
The authors wish to acknowledge Miss Kexin Chen from the CSE Department, CUHK, for the discussion of FSM-related content, and Miss Qijia Wang from the IE Department, CUHK, for her assistance in filming the demo video.
% \bibliography{References}

\begin{thebibliography}{10}
\providecommand{\url}[1]{#1}
\csname url@samestyle\endcsname
\providecommand{\newblock}{\relax}
\providecommand{\bibinfo}[2]{#2}
\providecommand{\BIBentrySTDinterwordspacing}{\spaceskip=0pt\relax}
\providecommand{\BIBentryALTinterwordstretchfactor}{4}
\providecommand{\BIBentryALTinterwordspacing}{\spaceskip=\fontdimen2\font plus
\BIBentryALTinterwordstretchfactor\fontdimen3\font minus \fontdimen4\font\relax}
\providecommand{\BIBforeignlanguage}[2]{{%
\expandafter\ifx\csname l@#1\endcsname\relax
\typeout{** WARNING: IEEEtran.bst: No hyphenation pattern has been}%
\typeout{** loaded for the language `#1'. Using the pattern for}%
\typeout{** the default language instead.}%
\else
\language=\csname l@#1\endcsname
\fi
#2}}
\providecommand{\BIBdecl}{\relax}
\BIBdecl

\bibitem{lee2023task}
S.~Lee, S.~Lee, Y.~Choi, J.~Ben-Othman, L.~Mokdad, K.-i. Hwang, and H.~Kim, ``Task-oriented surveillance framework for virtual emotion informatics in polygon spaces,'' \emph{IEEE Wireless Communications}, vol.~30, no.~3, pp. 104--111, 2023.

\bibitem{shi2023task}
Y.~Shi, Y.~Zhou, D.~Wen, Y.~Wu, C.~Jiang, and K.~B. Letaief, ``Task-oriented communications for 6g: Vision, principles, and technologies,'' \emph{IEEE Wireless Communications}, vol.~30, no.~3, pp. 78--85, 2023.

\bibitem{xu2023guest}
W.~Xu, Z.~Yang, D.~W.~K. Ng, O.~A. Dobre, L.-C. Wang, and R.~Schober, ``Guest editorial: Task-oriented communications for future wireless networks,'' \emph{IEEE Wireless Communications}, vol.~30, no.~3, pp. 16--17, 2023.

\bibitem{webb2023emergent}
T.~Webb, K.~J. Holyoak, and H.~Lu, ``Emergent analogical reasoning in large language models,'' \emph{Nature Human Behaviour}, vol.~7, no.~9, pp. 1526--1541, 2023.

\bibitem{kaplan2019siri}
A.~Kaplan and M.~Haenlein, ``Siri, \textsc{S}iri, in my hand: Who’s the fairest in the land? \textsc{O}n the interpretations, illustrations, and implications of artificial intelligence,'' \emph{Business horizons}, vol.~62, no.~1, pp. 15--25, 2019.

\bibitem{goertzel2007artificial}
B.~Goertzel and C.~Pennachin, \emph{{Artificial General Intelligence}}.\hskip 1em plus 0.5em minus 0.4em\relax Springer, 2007, vol.~2.

\bibitem{asay_2023_are}
\BIBentryALTinterwordspacing
M.~Asay, ``Are large language models wrong for coding?'' InfoWorld, 05 2023. [Online]. Available: \url{https://www.infoworld.com/article/3697272/are-large-language-models-wrong-for-coding.html}
\BIBentrySTDinterwordspacing

\bibitem{fsm}
P.~Gladyshev, ``Finite state machine approach to digital event reconstruction,'' \emph{Digital Investigation}, vol.~1, 05 2004.

\bibitem{Text2vec}
M.~Xu, ``Text2vec: Text to vector toolkit,'' \url{https://github.com/shibing624/text2vec}, 2023.

\bibitem{jocher_2020_yolov8}
\BIBentryALTinterwordspacing
G.~Jocher, ``\textsc{YOLO}v8 documentation,'' docs.ultralytics.com, 05 2020. [Online]. Available: \url{https://docs.ultralytics.com/}
\BIBentrySTDinterwordspacing

\bibitem{recognition}
\BIBentryALTinterwordspacing
ageitgey, ``ageitgey/face\_recognition,'' GitHub, 06 2019. [Online]. Available: \url{https://github.com/ageitgey/face\_recognition/}
\BIBentrySTDinterwordspacing

\bibitem{song2023llm}
C.~H. Song, J.~Wu, C.~Washington, B.~M. Sadler, W.-L. Chao, and Y.~Su, ``\textsc{LLM}-planner: Few-shot grounded planning for embodied agents with large language models,'' in \emph{Proceedings of the IEEE/CVF International Conference on Computer Vision}, 2023, pp. 2998--3009.

\bibitem{xia2023towards}
Y.~Xia, M.~Shenoy, N.~Jazdi, and M.~Weyrich, ``Towards autonomous system: \textsc{F}lexible modular production system enhanced with large language model agents,'' \emph{arXiv preprint arXiv:2304.14721}, 2023.

\end{thebibliography}

\end{document}